# Introduction : Why and How to Probe the Zone of Avoidance ?

Ofer Lahav

*Institute of Astronomy, Madingley Road, Cambridge CB3 0HA, UK*

**Abstract.** The motivation and major ways for probing the Zone of Avoidance (ZOA) are reviewed. Galaxies hidden behind the ZOA may have important implications for the internal dynamics of the Local Group, for the origin of its motion relative to the Microwave Background, and for the connectivity of the large scale structure. Current direct ('observational') methods for exploring the ZOA include eye-balling of plates, source identification in the IRAS data base, and pointed and blind-search observations in 21 cm. Interesting regions identified so far include the two crossing points of the Supergalactic Plane by the Galactic Plane (at Galactic longitude $l \sim 135°$, near Perseus-Pisces, and $l \sim 315°$, near the Great Attractor), the Puppis cluster (at $l \sim 240°, cz \sim 1500$ km/sec) and the Ophiuchus cluster (at $l \sim 0°, cz \sim 8400$ km/sec). New promising wavelengths are the $2\mu$ and the X-ray band. Indirect ('theoretical') approaches include 'Wiener reconstruction' from incomplete and noisy data, and using the peculiar velocity field as a probe of the mass distribution hidden behind the ZOA. The problem of source confusion at low Galactic latitude can be addressed by novel statistical methods, e.g. Artificial Neural Networks.

## 1. Introduction

Our own Galaxy is a major obstacle in probing the extragalactic sky. Local stars and dust block our view in most wavelengths, or cause confusion in identifying galaxies. If the 'Zone of Avoidance' (ZOA) covers Galactic latitudes $|b| < b_c$ then a solid angle $\omega = 4\pi \sin b_c$ is obscured. This corresponds typically to a fraction $\sim 20\%$ of the sky in the optical band and $\sim 10\%$ in the infrared.

As indicated by the term ZOA, the common view is that extragalactic studies should avoid this complicated region. Indeed, for some studies, e.g. for calculating correlation functions, it is not essential and even better to analyse regions of high Galactic latitude, as long as the sample is a 'fair sample'. However, for understanding other issues of the large scale structure it is crucial to unveil the whole-sky galaxy distribution. An earlier review (Kraan-Korteweg 1993) summarized previous studies, and the history of the subject is reviewed by Kerr in this volume. Here we shall attempt to summarize present and future methods for exploring the ZOA, and to show that the ZOA is not actually such an 'obscured' topic. In this meeting, we mainly refer to the ZOA in the context of the study of galaxy distributions. More generally, the ZOA also affects other

Figure 1. An equal area projection of the 2 hemispheres in Equatorial coordinates of galaxies larger than 1 arcmin from the UGC, ESO and MCG catalogues. The most prominent filament across the plot is the Supergalactic Plane. The empty bands are due to obscuration by the Milky way (after Lynden-Bell & Lahav 1988).

extragalactic measurements, e.g. the Cosmic Microwave Background and the X-ray background.

## 2. Implications of hidden matter behind the Milky Way

### 2.1. The connectivity of structure

Figure ?? shows the two hemispheres (in Equatorial coordinates) plotted in equal-area projection. The galaxies shown are from the UGC, ESO and MCG catalogues. The two empty bands are the ZOA. The Supergalactic Plane is seen in its full glory, but its full extent is hidden by the ZOA. There are also several other filaments running perpendicular to the ZOA (e.g. in the Southern hemisphere near Hydra, dubbed together as the 'dinosaur's foot' by Donald Lynden-Bell). This figure indicates that if we wish to know the extent of the Supergalactic Plane (and to learn if it is a 'plane' at all) we do need to unveil the structure behind the ZOA.

As pointed out by Kraan-Korteweg (1993), 5 out of the 8 apparent brightest galaxies lie in the ZOA : CenA, IC342, Maffei I and II and N4945. Since galaxies

## 2.2. The internal dynamics of the Local Group

It is intriguing to speculate on the existence of a nearby $M31$-like galaxy behind ZOA. Such a nearby galaxy, or several galaxies, would change dramatically our understanding of the dynamics of the Local Group (LG), currently believed to be dominated by the Milky Way and Andromeda. Discovering new members of the Local Group would also affect the deduction of the mass and the age of the Local Group from its dynamics (Kahn & Woltjer 1959; Lynden-Bell 1982) and the reconstruction of LG galaxy motions (Peebles 1990).

## 2.3. The origin of motion of the Local Group

Understanding the origin of motion of the Local Group relative to the Cosmic Microwave Background (CMB) radiation is of major importance for verifying the gravitational instability picture and for the deduction of the density parameter $\Omega_0$ and 'biasing' of the galaxies relative to the underlying mass distribution.

The CMB dipole is very accurately measured and is commonly interpreted as being due to the motion of the sun relative to the CMB radiation (at 370 km/sec towards $l = 264°, b = 48°$; Kogut et al. 1993). This motion is converted to the motion of the Local Group relative to the CMB (at about 600 km/sec towards $l = 268°, b = 27°$) by taking into account the rotation of the sun around the Galaxy and the relative motions of the Galaxy and Andromeda (see e.g. Yahil, Tammann & Sandage 1977; Lynden-Bell & Lahav 1988). Even if no nearby $M31$-like galaxy is to be found behind the ZOA, and the above calculation holds, then one needs to understand the sources of the motion of the Local Group at 600 km/sec. Some of the sources are probably behind the ZOA.

The expected peculiar velocity is calculated, assuming linear theory (Peebles 1980) by summing up the contribution from masses $M_i$ represented by the galaxies

$$\mathbf{v} \propto \frac{\Omega_0^{0.6}}{b} \sum_i \frac{M_i}{r_i^2} \hat{\mathbf{r}}_\mathbf{i} \;, \qquad (1)$$

where $\Omega_0$ is the density parameter and $b$ is the bias parameter (reflecting that light may not be a perfect tracer of the mass). Commonly the masses are assumed to be either equal or proportional to the galaxy luminosities (hence only the observed flux is required). The dipole in the distribution of IRAS and optical galaxies lies within $10° - 20°$ of the CMB dipole (e.g. Lynden-Bell, Lahav & Burstein 1989, Strauss et al . 1992), but the convergence of the dipole with distance is still an open question. In part this uncertainty is due to shot-noise, the finite depth of the samples, and redshift distortion, but another factor may well be the missing data behind the ZOA.

It is curious that the two major superclusters on opposite directions on the sky, the Great Attractor and Perseus-Pisces, which are playing the role in the 'tug of war' on the Local Group, are near the ZOA. Moreover, two major galaxies, CenA and IC342, are also in these opposite directions near the ZOA. Another example of a possible contribution to the Local Group motion is due the Puppis cluster at $l \sim 240°; \; b \sim 0°; \; cz \approx 1500$ km/sec. By comparison with the 2Jy

IRAS survey, it has been suggested (Lahav et al. 1993) that Puppis (which lies below the Supergalactic Plane) may contribute at least 30 km/sec to the motion of the Local Group *perpendicular* to the Supergalactic Plane ($V_{SGZ} \approx -370$ km/sec). Together with the Local Void (above the Supergalactic Plane, also crossed by the ZOA) and Fornax and Eridanus (below it) this may explain of the origin of the so-called "Local Anomaly".

## 3. Direct 'Observational' methods

### 3.1. Understanding the Milky Way and Extinction

The first step in exploring the extragalactic sky behind the ZOA is to understand our own Galaxy. The extinction and confusion with stars become more and more difficult as one approaches the Galactic Plane. Even the detailed extinction map of Burstein & Heiles (1982) only covers $|b| > 10^o$. New studies of the Galactic HI distribution (see Burton in this volume) are most useful to predict extinction, in particular if combined with the IRAS 100 $\mu$ emission maps and other Galactic measurements. The next step is to understand how Galactic extinction affects galaxy magnitudes and diameters. Cameron (1990) showed how to obscure high-latitude galaxies artificially, an approach which needs to be further developed.

### 3.2. Eye-balling of plates and optical redshift surveys

Earlier searches for galaxies on plates were carried out by Böhm-Vitense (1956) and Weinberger (1980). Focardi et al. (1984) suggested that Perseus-Pisces is connected to A569, as was later supported by radio and optical redshift measurements (Hauschildt 1987, Chamaraux et al. 1990). The region of Puppis was searched by Saito et al. (1990, 1991) on infrared film copies. Kraan-Korteweg (1993; also in this volume) carried out a systematic surveys on IIIaJ film copies of the ESO/SRC-survey in the directions of Hydra, and Hau surveyed Palomar red plates in the region centred on $l \sim 135^o$, where the Supergalactic plane is crossed by the GP.

The above are only some examples of many elaborate surveys, revealing thousands of new galaxies. They also provide target lists for follow-up of optical and radio redshift measurements (e.g. Fairall and Huchra in this volume), which then reveal the 3-dimensional structure behind the ZOA. However, the different searches are far from being on equal footing, and their interpretation is subject to selection effects, e.g. due to the plate material used, and the poorly known extinction at low latitudes.

The elaborate procedure of eye-balling calls for new automatic methods which will allow separating stars from galaxies in deep samples such as the APM (Maddox et al. 1990) and the $2\mu$m survey (Mamon in this volume). Another approach (Odewahn et al. 1991) utilizes Artificial Neural Networks which are 'trained' to separate galaxies from stars in a non-linear way (the method can also be used to classify galaxies into morphological types; Storrie-Lombardi et al. 1992). Such methods allow reproducibility and objective classification.

Figure 2. HI spectra from the Dwingeloo survey. The galaxy spectrum at the bottom is of NGC3359 at high Galactic latitude ($l = 144°, b = 49°$), while the spectrum at the top is of a galaxy behind the ZOA ($l = 95°, b = 2°$; number 34 in Henning 1992). (Burton, Ferguson, Hau, Henning, Kraan-Korteweg, Lahav, Loan & Lynden-Bell; work in progress.)

### 3.3. 21 cm

The search for galaxies behind the GP in the 21 cm line of neutral hydrogen (HI), was pioneered by Kerr & Henning (1987). They carried out (using the Green Bank 91m radio-telescope) blind observations towards 1900 points and detected 16 new galaxies. The spectra of the galaxies detected did look similar to those of galaxies at high Galactic latitude, indicating that the ZOA is almost transparent in HI measurements. Some show the typical spectral 'double horned' signature of a spiral galaxy. Others may well represent other populations, e.g. of dwarf galaxies and low surface brightness galaxies. This work has been extended by Henning (1992; and in this volume) and by other groups.

A new survey began recently using the Dwingeloo 25-m telescope in the Netherlands. This radio-telescope, recently used to map the Milky Way in HI (see Burton in this volume) is now dedicated to do blind search and pointed observations of galaxies out to 4000 km/sec as a collaboration of Cambridge, Dwingeloo, Groningen and Leiden. Figure ?? shows examples of galaxy spectra from this survey at high and low Galactic latitude, illustrating that galaxies can be detected behind the Galactic plane. A deeper new multi-beam whole-sky survey is planned using the Parkes 64-m and the Lovell 76-m (see Stewart in this volume.)

An alternative to the blind search approach is to have pointed 21 cm observations towards galaxy candidates selected from the IRAS Point Source Catalog or from optical plates. The two approaches are complementary. The blind search is systematic but rather slow, and includes population of dwarf and low surface brightness galaxies. The IRAS-selected HI search is biased towards spiral galax-ies, and is limited at very low Galaxy latitudes where IRAS is confusion limited. The selection effects in HI searches are the cutoff in redshift, the velocity resolution, and problems of solar interference.

### 3.4. IRAS

The IRAS Point-Source Catalog (PSC) has been exploited in recent years to provide galaxy candidates behind the ZOA (e.g. Lu et al. 1990, Yamada et al. 1993, and others in this volume). Different authors proposed different colour selection criteria to pick up galaxies from the PSC. These candidates were then followed up by HI radio surveys or by inspection of plates. The advantage of this approach is in producing a uniform sample, which can be related to the rest of the sky. Possible problems are confusion with Galactic sources and bias towards spiral galaxies.

### 3.5. 2 $\mu$m

This new wavelength for extragalactic surveys is promising for two reasons: (i) it probes the old 'stable' stellar population in galaxies and (ii) it little suffers from Galactic extinction. On the other hand, it suffers confusion with Galactic sources at low Galactic latitude.

Two major surveys are now carried out in $2\mu$m. The 2 Micron All-Sky Survey (2MASS), described by Huchra in this volume, which will cover the sky at 3 bands. This survey will yield a sample of about 100000 galaxies down to 13-14 $K$ magnitudes within $10^o$ of the GP. The second survey, DENIS, described in this volume by Mamon, will cover the Southern sky.

### 3.6. X-ray

This wavelength is discussed by Fabian in these Proceedings. X-ray selected clusters are more easily detected than optical clusters at low Galactic latitude (e.g. Lahav et al. 1989). The whole-sky HEAO1 and Rosat surveys are in particular useful for the ZOA problem.

## 4. Indirect 'theoretical' methods

### 4.1. Wiener reconstruction of all sky surveys

Previous corrections for the unobserved ZOA (e.g. for the dipole calculation) were done, in a somewhat ad-hoc fashion, by populating the ZOA uniformly according to the mean density, or by 'shifting' or interpolating the structure below and above the Galactic Plane to create 'mock sky' in the ZOA (e.g. Lynden-Bell et al.1989, Strauss et al. 1992, Hudson 1993). An alternative approach allows mask inversion by regularization with a 'Wiener filter' (the ratio of signal to signal+noise), in the framework of Bayesian statistics and Gaussian random fields, by assuming a prior model for the power-spectrum (or correlation function) of the galaxies. Lahav et al. (1994) have applied this method, utilizing spherical harmonics, to the projected IRAS 1.2Jy survey. Their whole-sky noise-free reconstruction confirms the connectivity of the Supergalactic plane across the ZOA (at $l \sim 135^0$ and $l \sim 315^o$) and the Puppis cluster at $l \sim 240^0$. Detailed discussion of this method and pictorial reconstructions appear elsewhere in this volume (see contributions by Hoffman and Lahav).

### 4.2. Dynamical reconstruction of the mass behind the ZOA

Instead of looking at the distribution of galaxies at the ZOA, one can use the peculiar velocity field on both sides the ZOA to predict the *mass*-density distribution behind the Galactic Plan. The Potent method (Dekel, Bertschinger & Faber 1990) allows to recover the potential from line-of-sight peculiar velocities and hence the mass-density field. Kolatt, Dekel & Lahav (1994) have used the method to look in the direction of the ZOA.

The main dynamical features found at a distance $r \sim 4000$ km/sec are (a) the peak of the Great Attractor connecting Centaurus and Pavo at $l \sim 330°$, (b) a moderate bridge connecting Perseus-Pisces and Cepheus at $l \sim 140°$, and (c) an extension of a large void from the southern Galactic hemisphere into the ZOA near the direction of Puppis, $l \sim 220° - 270°$. They found a strong correlation between the mass density and the IRAS and optical galaxy density at $b = \pm 20°$, which indicates that the main dynamical features in the ZOA should also be seen in galaxy surveys through the Galactic Plane. Note that at 4000 km/sec the Potent resolution of 1200 km/sec corresponds to $17°$, so one can only reconstruct the gross features across the ZOA, rather than the individual clusters. The gravitational acceleration at the Local Group, based on the *mass* distribution out to $\sim 6000$ km/sec, is found to be strongly affected by the mass distribution in the ZOA: its direction changes by $31°$ when the $|b| < 20°$ ZOA is included, bringing it to within $4° \pm 19°$ of the CMB dipole.

## 5. New individual structures behind the ZOA

Various structures behind the GP are discussed by different authors in detail in this volume. Here we only point out some examples of interesting regions.

### 5.1. The Great Attractor and A3627

From the study of the peculiar velocity field of elliptical galaxies Lynden-Bell et al. (1988) predicted, using a simple spherical model, that the centre of the 'Great Attractor' is at $l = 307°, b = 9°$, i.e. behind the ZOA. It is interesting that several recent studies also suggest similar directions. Kraan-Korteweg & Woudt (these Proceedings) have studied this region by visual inspection of IIIaJ copies of the ESO/SRC sky surveys and redshift measurements. They find a peak centred on the rich ACO cluster A3627 ($l = 325°, b = -7°, cz = 4300$ km/sec). Reconstruction methods, although with smoothing larger than cluster-scale, also suggest an enhancement in this neighbourhood. Kolatt et al. (1994) find a peak in *mass-density* in the Potent reconstruction at ($l = 320°, b = 0°, cz = 4000$ km/sec) and Hoffman (this volume) also finds a peak in the galaxy distribution in that region.

### 5.2. Puppis

This cluster was recognized independently by Kraan-Korteweg & Huchtmeier (1992), Yamada et al. (1992), and Scharf et al. (1992). To quantify the impor-tance of Puppis relative to other structures in the local universe Lahav et al. (1993) supplemented the IRAS 2 Jy $|b| > 5^o$ redshift survey (Strauss et al. 1992) with an IRAS-selected sample in the direction of Puppis ($|b| < 5^o; 230^o < l < 260^o$), which consists of 32 identified galaxies, 12 of them with measured redshift. It was found that the projected number counts of galaxies brighter than 2 Jy in Puppis is about half that of Virgo. A Spherical Harmonic reconstruction (Figure ??) shows that out to a distance of 3000 km/sec Puppis is second only to Virgo. However, Puppis does not seem to be a virialized cluster, and it is not seen in the HEAO1 and ROSAT X-ray maps (K. Jahoda and H. Böhringer, private communication)

Figure 3. Spherical Harmonic reconstruction with coefficients up to $l_{max} = 10$, of galaxies with 500 km/sec $< cz_{LG} <$ 3000 km/sec in the 2Jy IRAS ($|b| > 5^o$) combined with our Puppis sample ($|b| < 5^o$). Plots are equal area hemispheres with the left-hand side plot centred on Galactic $l = 240°$, $b = 0°$. The Galactic Plane runs horizontally across the plots (solid line), dashed lines bound the region $|b| < 5°$ and longitudes are indicated. The South Galactic hemisphere is at the top. The lightest solid contour is at the mean, dashed and solid contours indicate densities below and above the mean respectively. Contour separation is 3 times the shot-noise level. Associations with local structures are labelled: V - Virgo, PU - Puppis, F - Fornax E - Eridanus, Leo - Leo, Cen - Centaurus, PIT - Pavo-Indus-Telescopium, Cam - Camelopardalis and UM - Ursa Major. (From Lahav et al. 1993).

### 5.3. Ophiuchus

This cluster, behind the Galactic Centre (!), has been studied by Wakamatsu (this volume, and references therein). His catalogue of 6000 galaxies suggests that there is a big supercluster around the Ophiuchus cluster ($l = 0^\circ, b = 8^\circ, cz \approx$ 8500 km/sec). This is also one of the brightest known X-ray clusters (e.g. Wakamatsu & Malkan 1980, Lahav et al. 1989 and Fabian in this volume). Moreover, the Ophiuchus supercluster seems to be connected to the Hercules supercluster. Djorgovski et al. (1990) speculated that Ophiuchus is also connected to the Sagittarius cluster (at $l = 359^\circ, b = 8^\circ, cz \approx 8600$ km/sec).

## 6. Discussion

This is the first meeting dedicated to the ZOA problem, bringing together researchers who are using different methods and working at different wavelengths. The following are some of the important issues calling for discussion:

- How to produce a combined Galactic extinction map based on HI, IRAS etc. ?

- Who is observing what and where in the ZOA ?

- What are the selection effects in optical, IRAS, 2$\mu$m, 21 cm and X-ray surveys ?

- How to combine different surveys to produce a uniformly sampled map ?

- How to implement algorithms for automatic separation of galaxies from stars at low Galactic latitudes ?

**Acknowledgments.** I am grateful to H. Ferguson, R. Kraan-Korteweg, A. Loan and D. Lynden-Bell for their comments on this manuscript, and to my collaborators to some of the work described here for their contribution and helpful discussions.

**Discussion**

*T. Yamada*: You said that one of the motives of investigating the galaxy distribution behind the ZOA is to construct a '3-D fair sample'. What do you mean by 'fair sample' in this case ? If you see deeper universe (as deep as you need!), you will get a 'fair sample' at high latitude and you might not have to study the structure behind the Milky Way ?

*O. Lahav*: The definition of a 'fair sample' depends of course on the statistic of interest. If we want to study nearby filaments, the structure behind the ZOA is important. Furthermore, if we go very deep we may encounter new problems, e.g. evolution.

*C. Balkowski*: Could you tell more about the new method for star/galaxy separation ?

*O. Lahav*: Computer algorithms called Artificial Neural Networks 'learn' (essentially by least square minimization with respect to free parameters called 'weights') to classify images into stars and galaxies from a set of examples for which the answer is already known (e.g. from a human expert). After a network has been 'trained' (so the weights were fixed) new (previously unclassified) objects are presented to the network. The network provides probabilities for an object being a star or a galaxy.

*L. Gouguenheim*: I have an additional remark about selection effects - the majority of neutral hydrogen studies starting from IRAS sources, selected from their IR colours, have still focused on optically identified objects, because of the better detection rate. Galaxies are then found in region of low obscuration.

*O. Lahav*: Blind searches in HI may overcome some of these problems.